\documentclass[reprint]{raa}           
\usepackage{graphicx,times}
\usepackage{natbib}
\usepackage{amssymb,amsmath}
\bibpunct{(}{)}{;}{a}{}{,}

\usepackage[a4paper=true,driverfallback=dvipdfm,pagebackref=true]{hyperref}
\hypersetup{pdftitle = The title of my PDF, pdfauthor = My name, pdfsubject= The subject, pdfkeywords = keyword1 keyword2 keyword3} 
\hypersetup{colorlinks = true, linkcolor = green, anchorcolor = red, citecolor = blue, filecolor = red, pagecolor = red, urlcolor = red}

\begin{document}

   \title{Searching for $\textbf{\textit{r}}$-process-enhanced stars in the LAMOST Survey I: The Method
$^*$
\footnotetext{\small $*$ Supported by the National Natural Science Foundation of China.}
}

 \volnopage{ {\bf 20XX} Vol.\ {\bf X} No. {\bf XX}, 000--000}
   \setcounter{page}{1}

   \author{Tian-Yi Chen\inst{1,2}, Jianrong Shi\inst{1,2},  Timothy C. Beers\inst{3},  Hongliang Yan\inst{1,2}, Qi Gao\inst{1,2}, Chun-Qian Li\inst{1,2}, Hai-Ning Li\inst{1}, Gang Zhao\inst{1,2}
   }

\institute{CAS Key Laboratory of Optical Astronomy, National Astronomical Observatories, Chinese Academy of Sciences, Beijing 100101, China; {\it sjr@nao.cas.cn}\\
 \and     School of Astronomy and Space Science, University of Chinese Academy of Sciences, Beijing 100049, China\\ 
\and    Department of Physics and JINA Center for the Evolution of the Elements, University of Notre Dame, Notre Dame, IN 46556, USA\\
\vs \no
  {\small Received 2020 Month Day; accepted 2020 Month Day}
}

\abstract{The abundance patterns of $r$-process-enhanced stars contain key information required to constrain the astrophysical site(s) of $r$-process nucleosynthesis, and to deepen our understanding of the chemical evolution of our Galaxy. In order to expand the sample of known $r$-process-enhanced stars, we have developed a method to search for candidates in the LAMOST medium-resolution ($R \sim 7500$) spectroscopic survey by matching the observed spectra to synthetic templates around the Eu II line at 6645.1\,\AA. We obtain a sample of 13 metal-poor ($-2.35<{\rm [Fe/H]}<-0.91$) candidates from 12,209 unique stars with 32,774 medium-resolution spectra. These candidates will be further studied by high-resolution follow-up observations in the near future. Extensions of this effort to include larger samples of stars, in particular at lower metallicity, using the strength of the Ba II line at 6496.9\,\AA, are described. 
} 
\authorrunning{T.-Y Chen et al. }            
\titlerunning{Searching for $\textbf{\textit{r}}$-process-enhanced stars}  
\maketitle

%
\section{Introduction}           
\label{sect:intro}

\hspace{15pt}Elements heavier than the iron peak are produced by neutron-capture processes, roughly half of which result from the rapid neutron-capture process ($r$-process). Although the basic concepts on $r$-process production were first proposed over six decades ago \citep{Burbidge1957, Cameron1957}, its astrophysical sites have remained a long-standing question.

Recently, the neutron star merger (NSM) paradigm \citep{Lattimer1974, Rosswog2014,Thielemann2017} has received strong support from observation of the gravitational wave event GW\,170817 by LIGO/Virgo \citep{ Abbott2017}, the photometric and spectroscopic follow-up of its electromagnetic counterpart, the kilonova SSS17A \citep{Drout2017, Kilpatrick2017, Shappee2017}, provided the first direct evidence that NSMs are at least one site for $r$-process-element production \citep{Watson2019}. In addition, the discovery of extremely $r$-process-enhanced (RPE) stars in the ultra-faint dwarf galaxy Reticulum II \citep{Ji2016, Ji2019, Roederer2016} , as well as moderately RPE stars in the ultra-faint dwarf Tucana III \citep{Hansen2017, Marshall2019} provided additional support for this hypothesis.

However, the frequency of NSMs, the total amount of $r$-process-elements produced by them in the Galaxy, and the predicted ranges in their yields suggest that NSMs may not be the sole source. Other possible scenarios still under consideration include jets in magneto-rotational supernovae \citep[Jet-SNe,][]{Cameron2003, Winteler2012, Nishimura2015, Moesta2018}, core-collapse supernovae \citep[CCSNe,][]{Arcones2007, Wanajo2013,Thielemann2017} and collapsars \citep{Siegel2019}. These may lead to different abundance patterns (particularly between the $r$-process peaks) and different levels of enrichment.

The abundances of the $r$-process elements, which can be derived from observations in the atmospheres of  old metal-poor stars, carry the 
''fossil record" in their natal gas, polluted by previous enrichment events.  They can provide constraints on the conditions of the responsible nucleosynthesis sites, and have recorded important clues to the chemical evolution of our Galaxy. Several sub-classes of RPE stars, with various abundance patterns and relative levels of  enrichment, have been identified:  the $r$-I, $r$-II, and limited-$r$ stars. The $r$-I and $r$-II stars are moderately ($+$0.3$\le$[Eu/Fe]$\le+$1.0) and highly ([Eu/Fe]$>+$1.0) enhanced in heavy $r$-process elements ($Z\ge$ 56), respectively \citep{BeersChristlieb2005}, and require [Ba/Eu]$<$0 to ensure that the $r$-process dominates the $n$-capture process over the slow neutron-capture process ($s$-process). Variations in the enhancement levels of  the actinides (Th and U), compared to the second $r$-process-peak elements such as Eu, exist for a subset of RPE stars, known as 
''actinide boost" stars \citep{Hill2002, MashonkinaL2014, Holmbeck2018}. The limited-$r$ stars \citep{Frebel2018} exhibit low enrichment in the heavy $r$-process elements ([Eu/Fe]$<+$0.3) but higher abundances among elements in the first $r$-process peak, such as Sr, Y, and Zr; these are often identified by having [Sr/Ba]$>+$0.5.

Astronomers have been searching for RPE stars for decades. Earlier efforts, including the Hamburg/ESO Survey \citep[HES,][]{ Christlieb2004, Barklem2005} and spectroscopic studies of the Milky Way dwarf satellite galaxies \citep{Shetrone2003, Roederer2016, Hansen2017} and globular clusters \citep{Gratton2004, Sobeck2011}, have found tens of $r$-II stars and more than one hundred $r$-I stars in total. More recently, an extensive collaboration known as the $R$-Process Alliance (RPA) \citep{Hansen2018, Sakari2018, Ezzeddine2020, Holmbeck2020} has greatly accelerated the pace for RPE star discovery.

The Large Sky Area Multi-Object Fiber Spectroscopic Telescope (LAMOST) \citep{Zhao2006, Cui2012, Zhao2012} has been conducting the medium-resolution survey (MRS,  R$\simeq$ 7500) since September 2017, and obtained over one million spectra in DR6 and about four million spectra in DR7, respectively. Fortunately, the wavelength ranges of the LAMOST MRS spectra cover the absorption lines of Eu II and Ba II, which provides the opportunity to carry out a systematic search for RPE candidates in large numbers, and expand the current sample, after they have been confirmed with higher resolution spectroscopic follow-up.

In this work, we describe an initial effort to identify candidate RPE stars, by matching synthetic spectra for RPE stars to the LAMOST MRS spectra. The nature  of our stellar sample and the detailed selection method are presented in Sections \ref{sec:sample} and \ref{sec:method}, respectively. Validation of our method is provided in Section \ref{sec:validation}, and the estimated errors obtained for our first-pass abundance determinations are discussed in Section \ref{sec:error}. Section \ref{sec:sum} presents brief conclusions and a perspective on the next planned steps in this effort.


\section{Stellar sample} 
\label{sec:sample}
\hspace{15pt} The spectra used in this study are  a subset of the LAMOST MRS sample, conducted from September 2017 to April 2019 (DR6 and a portion of DR7). In order to obtain approximate abundances of Eu (and Ba) we require first-pass estimates of the stellar atmospheric parameters, hence in this paper we select stars with stellar parameters provided by the LAMOST Stellar Parameter Pipeline \citep[LASP,][]{Luo2015} from the low-resolution survey (LRS). 

The sample consists of 40,823 stars, spanning a wide range of stellar atmospheric parameters: 3500\,K $\leq$ {$\mathrm{\emph{T}_{eff}}$} $\leq$  7000\,K, 0.20 $\leq$ log \emph{g} $\leq$ 4.90, $-$2.50 $\leq$ [Fe/H] $\leq +0.70$. The majority of our targets were observed several times, and only those spectra with signal to noise ratio (S/N) higher than 20 have been selected for further investigation, so that reliable Eu and Ba abundance estimates could be obtained. 

\section{Methods} 
\label{sec:method}
\hspace{15pt}The method we adopted to estimate the Eu (and Ba) abundances is matching the observed spectra to synthetic templates with corresponding stellar parameters. This method is the basis of the LAMOST stellar parameter pipeline designed by the team at Peking University \citep[LSP3,][]{Xiang2015}, which is similar to that used by \cite{Gao2019}.

\subsection{Preprocessing of the observed spectra}

\hspace{15pt}Several steps have been taken to prepare an observed spectrum before the matching. After converting from the vacuum-wavelength to the air-wavelength scale, we performed a cross-correlation with the templates with corresponding stellar parameters 
to measure the radial-velocity 
($\mathrm{\emph{V}_{r}}$), and applied this correction to the blue and red arms of the MRS spectra. 

\subsection{The synthetic template spectra}

\hspace{15pt}To obtain the required templates for matching, we generated a series of synthetic spectra based on linear grids in stellar atmospheric-parameter space in advance. The template spectra were generated using the SPECTRUM synthesis code \citep[V2.76, 2010,][]{Gray1999} with the Kurucz ODFNEW atmospheric models \citep{Castelli2003}. In the calculation, the atomic line data for Eu II and Ba II are from \citet{Zhao2016}, and the standard Solar abundance is from \citet{Asplund2009}.

The stellar parameter ranges of our template grids are: 

3500\,K $\leq$ {$\mathrm{\emph{T}_{eff}}$} $\leq$  7000\,K in steps of 100\,K 

0.0 $\leq$ log \emph{g} $\leq$5.0 in steps of 0.25\,\ {dex} 
 
for $-$2.6 $\leq$ [Fe/H] $\leq-$1.0 in steps of 0.2\,\ {dex}, for $-$1.0$\leq$ [Fe/H] $\leq+$0.5 in steps of 0.1\,\ {dex}
 
for $-$0.4 $\leq$ [Eu/Fe] $\leq+1.0$ in steps of 0.1\,\ {dex},  for $+$1.0 $\leq$ [Eu/Fe] $\leq+$2.0 in steps of 0.2\,\ {dex}

The micro-turbulence was set according to \cite{Edvardsson1993}, and the resolution was set as 0.1\,\AA\ per pixel.

Given that the resolution of a LAMOST MRS spectrum is 0.1 $\sim$ 0.2\,\AA, and varies with different wavelengths and fibers, the template spectra are degraded to the observed resolution before comparison.

\subsection{Estimation of Europium abundances}
\hspace{15pt}The wavelength coverage of the LAMOST MRS spectra is about 6300 to 6800\,{\AA} for the red arm, and 4950 to 5350\,{\AA} for the blue arm. In the red arm, there is a Eu II line at 6645.1\,{\AA} available for abundance derivation; we set up a window from 6605 to 6685\,{\AA}  for both the observed and template spectra for Eu abundance determination.

\begin{figure}[!t]
\includegraphics[width=\textwidth, angle=0]{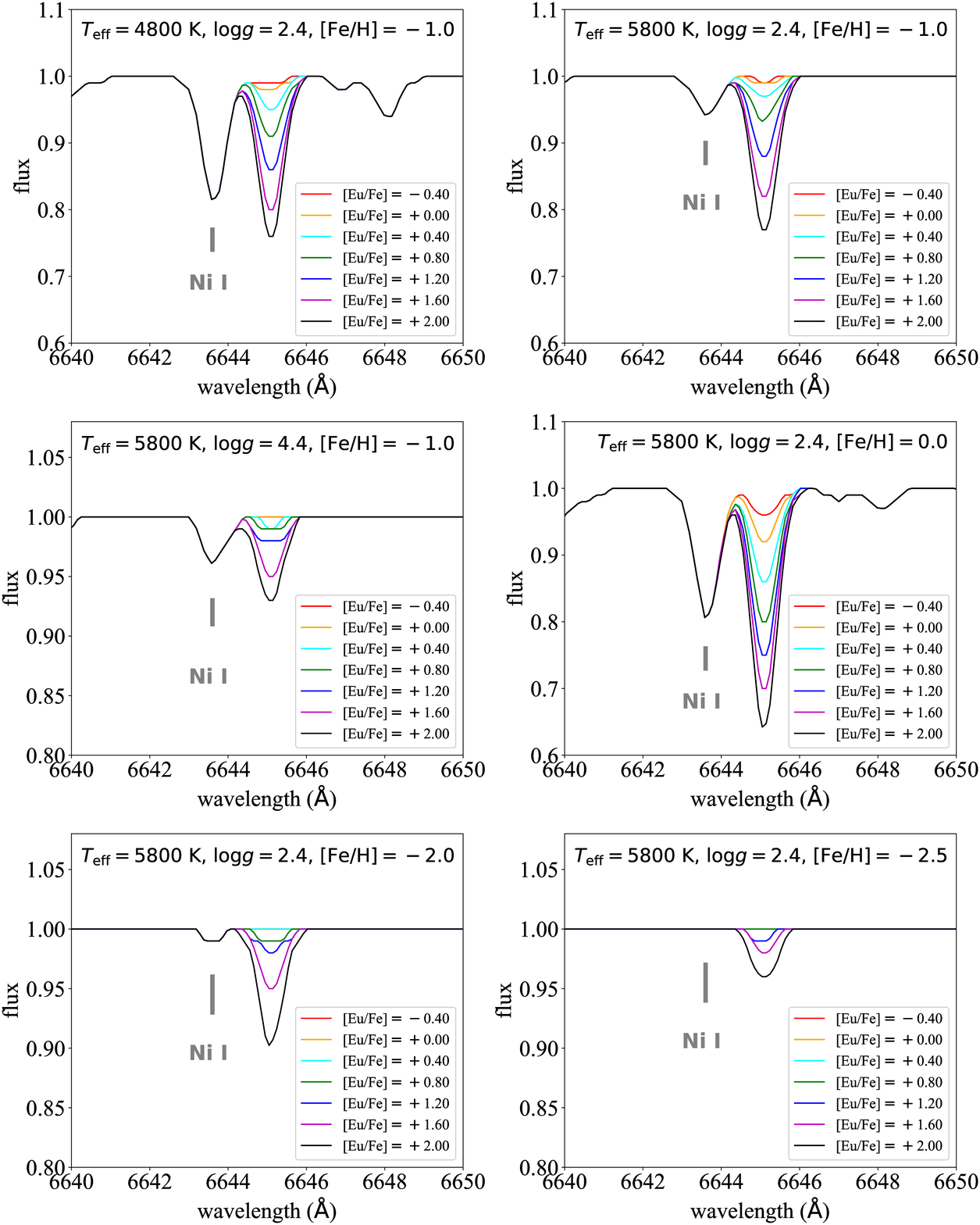}
\centering
\caption{Theoretical Eu II lines at 6645.1\,\AA\ with different stellar parameters, interpolated using template grids close to the adopted parameters. Eu II absorption lines with [Eu/Fe] varying from $-$0.40 to $+$2.00  are  shown with different colors. \label{fig:1}}
\end{figure}

For each observed LAMOST MRS spectrum, a set of templates have been generated by interpolating the template grids calculated beforehand, and adopting the stellar  parameters ($\mathrm{\emph{T}_{eff}}$,  log \emph{g} , [Fe/H]) provided by LASP (LRS) with [Eu/Fe] varying  from $-$0.40 to $+$2.0. Figure\ref{fig:1} presents six sets of template spectra for the listed stellar parameters, and shows how the Eu II line profiles change with different sets of stellar parameters (in different panels) and Eu abundances (in each panel).  

\begin{figure}[!t]
\includegraphics[width=\textwidth, angle=0]{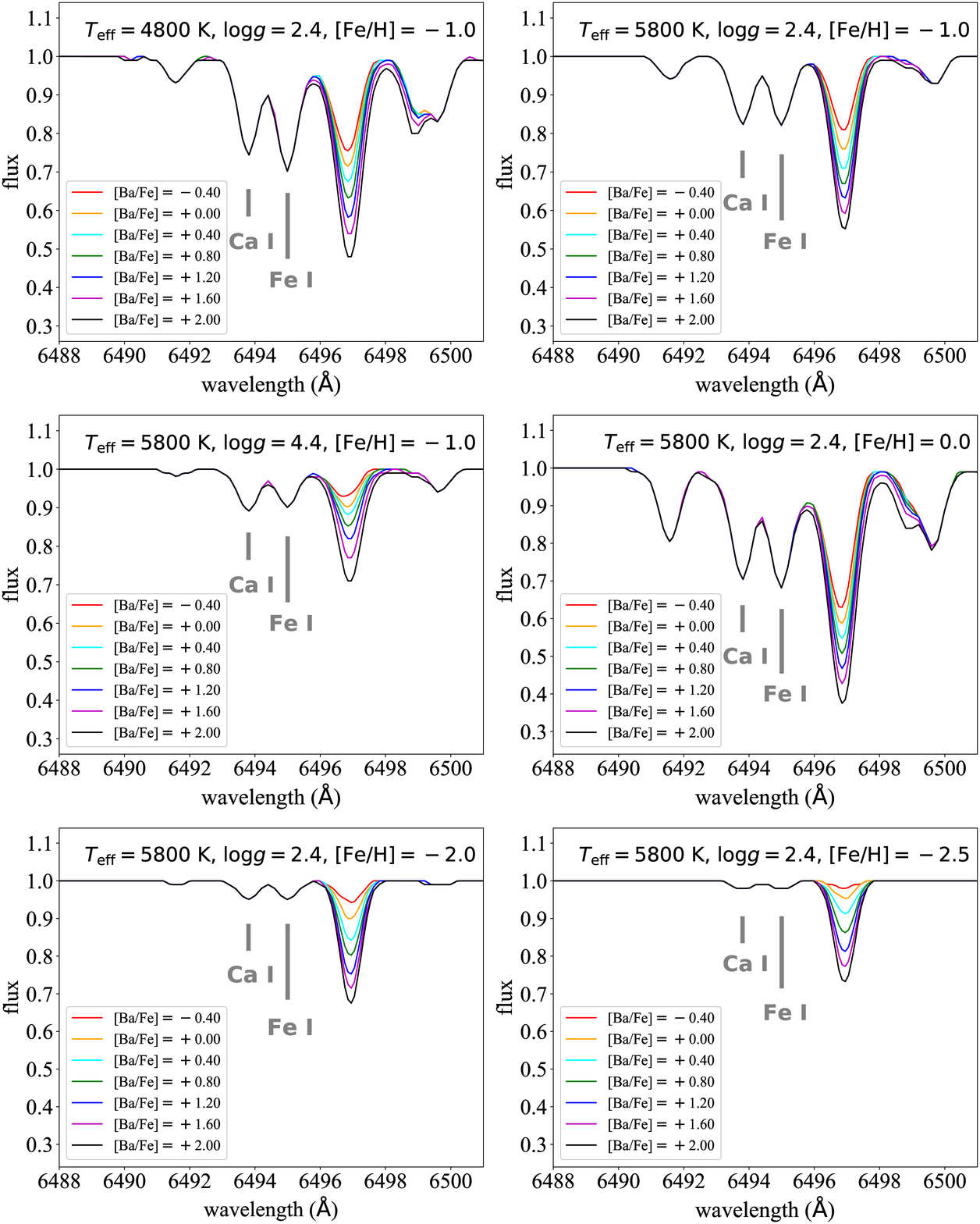}
\centering
\caption{Theoretical template spectra around the Ba II line at 6496.9\,\AA\ corresponding to six sets of stellar parameters. The different colors indicate the abundances of [Ba/Fe] varying from $-$0.40 to $+$2.00.} 
 \label{fig:2}
\end{figure}

To avoid the uncertainty introduced by continuum estimation, we instead scaled the synthetic templates with a third-order polynomial in order to match the flux level of the observed spectra, as was done by \cite{Xiang2015} and \cite{Gao2019}.

Then, the chi-squares ($\chi^2$) are calculated between these scaled templates of different Eu abundances and the observed spectra. The broadening due to the instrument has also been taken into consideration.  
The $\chi^2$ is defined as:
\[{\chi}^2=\sum_{i=1}^{N} \frac{(O_i - T_i)^2}{{\sigma}_i^2},\]
where ${O_i}$ and ${T_i}$ represent the fluxes of the ${i_{th}}$ point for the observed and the template spectrum, respectively, ${\sigma}_i$ is the error of the observed flux at the ${i_{th}}$ pixel, and $N$ is the number of pixels in the spectral window. 

The Eu abundances were derived with the corresponding minimum $\chi^2$ value by a third-order polynomial fit to the relation between [Eu/Fe] and $\chi^2$. Considering the $r$-enhanced criterion of [Ba/Eu]$<$0, we also estimated the Ba abundance  using the Ba II line at 6496.9\,\AA\ in the wavelength range of 6457 - 6537\,\AA\ in the same way. The synthetic template spectra around the Ba II line with the same six sets of stellar parameters are shown in Figure \ref{fig:2}. For stars with multiple LRS observations, only the stellar parameters from the highest S/N spectra were applied.

\subsection{Selection criteria}

\hspace{15pt}The Eu II line at 6645\,{\AA} is relatively weak, so it is rather sensitive to the noise level in the spectrum, which can lead to a large uncertainty in the estimation of the [Eu/Fe]  abundance. Therefore, we define three values to mitigate these effects: a) the depth of the Eu II line at 6645.1\,\AA\ (D); b) the average noise around the wavelength range of 6605-6685\,{\AA} (N); and, c) the standard deviation of the residuals between the observed spectrum and the best-matching template (S).

Only the stars with MRS spectra conforming to the following conditions were regarded as candidate RPE stars:

\begin{enumerate}
\item High $r$-process (Eu) abundance: [Eu/Fe]$\textgreater+$0.5,
\item Line is stronger than the noise: D $\textgreater$ N \& D $\textgreater$ S,
\item The $r$-process dominates the neutron-capture process: [Ba/Eu]$<$0.
\end{enumerate}

\begin{figure}[!t]
\includegraphics[width=\textwidth, angle=0]{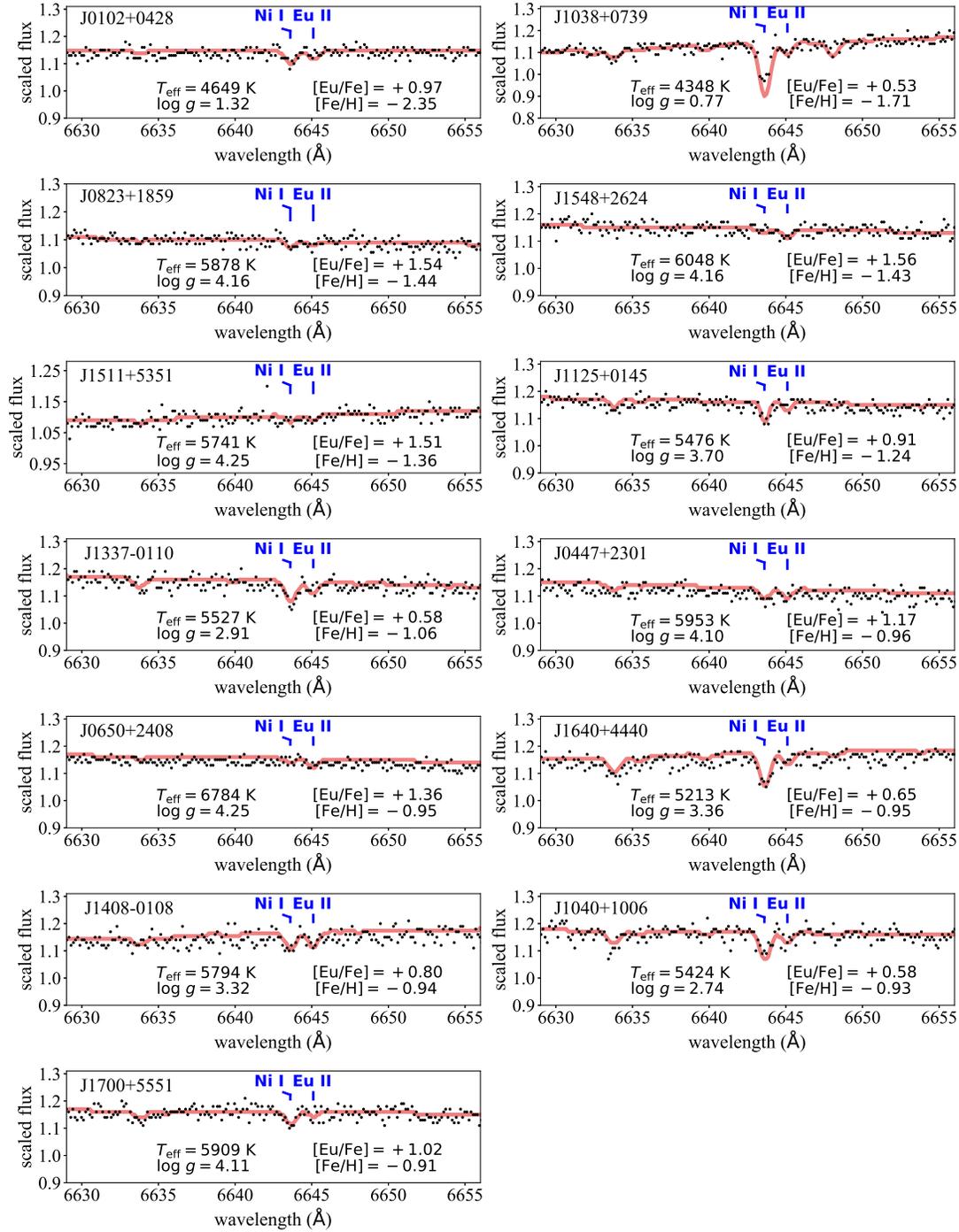}
\centering
\caption{13 metal-poor candidate RPE stars. The black dots are the LAMOST MRS spectra, while the red lines are the best-fitting templates. 
\label{fig:3}}
\end{figure}

We identified 13 metal-poor PRE candidates satisfying these criteria, and present them in Table\,\ref{tab:1}. Figure\,\ref{fig:3} presents their spectra and corresponding matching results; the Eu II line at 6645.1\,\AA\ can be clearly seen. 

 \begin{table}
\bc
\caption[]{List of the metal-poor RPE candidates
\label{tab:1}}
 \begin{tabular}{llrrrrrc}
  \hline\noalign{\smallskip}
  \hline
Star&Date& {$\mathrm{\emph{T}_{eff}}$} (K) & log \emph{g} & [Fe/H] & [Eu/Fe]&[Ba/Fe] & S/N \\
  \hline
J010212.66$+$042824.0&20170928&4649&1.32&$-$2.35&$+$0.97&$-$0.78&78.55\\
J044752.25$+$230109.8&20171029&5953&4.10&$-$0.96&$+$1.17&$-$0.44&69.13\\
J065034.18$+$240838.8&20171231&6784&4.25&$-$0.95&$+$1.36&$-$0.36&88.73\\
J082353.86$+$185934.5&20181128&5877&4.16&$-$1.44&$+$1.54&$-$0.13&56.97\\
J103848.45$+$073951.8&20171204&4348&0.77&$-$1.71&$+$0.53&$-$1.27&60.33\\
J104016.05$+$100635.7&20180305&5424&2.74&$-$0.93&$+$0.58&$-$0.11&53.29\\
J112501.85$+$014503.0&20171231&5476&3.70&$-$1.25&$+$0.91&$-$0.23&52.54\\
J133716.75$-$011000.6&20180202&5527&2.91&$-$1.06&$+$0.58&$-$0.21&59.01\\
J140816.00$-$010856.2&20180126&5794&3.32&$-$0.94&$+$0.80&$+$0.54&52.06\\
J151125.47$+$535118.5&20180525&5741&4.25&$-$1.36&$+$1.51&$-$0.52&76.49\\
J154846.00$+$262409.3&20180326&6048&4.16&$-$1.43&$+$1.56&$+$0.11&65.75\\
J164006.76$+$444010.4&20180531&5213&3.36&$-$0.95&$+$0.65&$-$0.39&54.12\\
J170018.61$+$555136.4&20180503&5909&4.11&$-$0.91&$+$1.02&$+$0.05&57.37\\
\hline
\end{tabular}
\ec
\end{table}

\section{Validation of the method}
\label{sec:validation}

\hspace{15pt}It is important to validate the results of our method. Some objects in our sample had already been identified as RPE stars by high-resolution (HR) spectroscopy, thus, we performed the estimation of [Eu/Fe] with the LAMOST MRS spectra for the objects which are in common with \citet{Sakari2018}, using their stellar parameters. The two stars in common were successfully selected as RPE candidates through our method; the derived Eu abundances have differences within  0.2\,\ {dex} of the HR analysis. 

\begin{figure}[!t]
\includegraphics[scale=0.5]{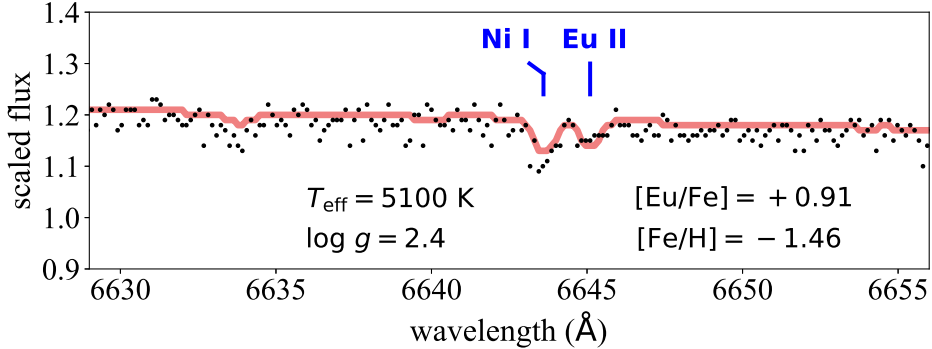}
\centering
\caption{Best-fit template (red line) to the scaled LAMOST MRS spectrum (black dots) of the Eu II line at 6645.1\,\AA\ for the previously known $r$-II star (RAVE J040618.2$-$030525). The corresponding Eu abundance is given in the panel. The adopted stellar parameters are {$\mathrm{\emph{T}_{eff}}$} = 5100\,K, log \emph{g} = 2.4, [Fe/H] = $-$1.46 from the HR analysis of  \citet{Rasmussen2020}, with a HR result of  [Eu/Fe] = $+$1.17. }
\label{fig:4}
\end{figure}

Figure~\ref{fig:4} provides an example of the fitting result for the previously known $r$-II star (RAVE J040618.2$-$030525). We adopted the stellar parameters of {$\mathrm{\emph{T}_{eff}}$} = 5100\,K, log \emph{g} = 2.4 , [Fe/H] = $-$1.48\ from \citet{Rasmussen2020}, and obtained an abundance of [Eu/Fe] as $+$0.91, which is consistent with the value of [Eu/Fe] $= +$1.17 from HR, considering the typical measurement uncertainty.

\section{Error estimation}
\label{sec:error}
\hspace{15pt}Many factors could result in uncertainties in the determination of [Eu/Fe]. Here we mainly discuss the random errors due to the quality of the observed spectra, and the uncertainties in the  stellar atmospheric parameters provided by LASP. 

\subsection{Errors due to the quality of spectra} 
\begin{figure}[!t]
\includegraphics[scale=0.5]{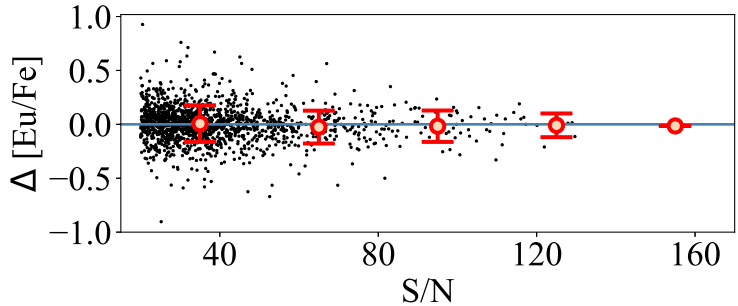}
\centering
\caption{Differences of individual [Eu/Fe] measurements from the mean of multiple visits versus the S/N of the corresponding spectra for the 670 stars with repeat observations among the RPE candidates. The black dots represent the individual measurements, while the red points represent averages (and their standard deviations) in bins of size 30 for S/N.}
\label{fig:5}
\end{figure}

\hspace{15pt}Taking advantage of the multiple visits of a target by the LAMOST MRS, we can estimate the random errors due to the quality of the observed spectra. Among the RPE candidates found by our template-matching method, 670 stars have multiple MRS spectra available. We calculated the Eu abundances for each individual spectrum, and derived the average Eu abundance for the repeated observations. The results are presented as individual offsets from the means ($\Delta$[Eu/Fe]) versus S/N of the single spectrum in Figure \ref{fig:5}, which indicates that $\Delta$[Eu/Fe] gradually declines from 0.2 to 0.1\,\ {dex} with S/N increasing from 20 to 150. Our results suggest that high S/N ($>$50) MRS spectra are needed for searching for RPE candidates. However, when the Eu II line is sufficiently strong (e.g., in the case of cooler $r$-II stars), candidates can be readily recognized with our method even from spectra with relatively low S/N ($\sim$20).

\subsection{Errors due to uncertainties in the stellar atmospheric parameters} 
\begin{figure}[!t]
\includegraphics[scale=0.5]{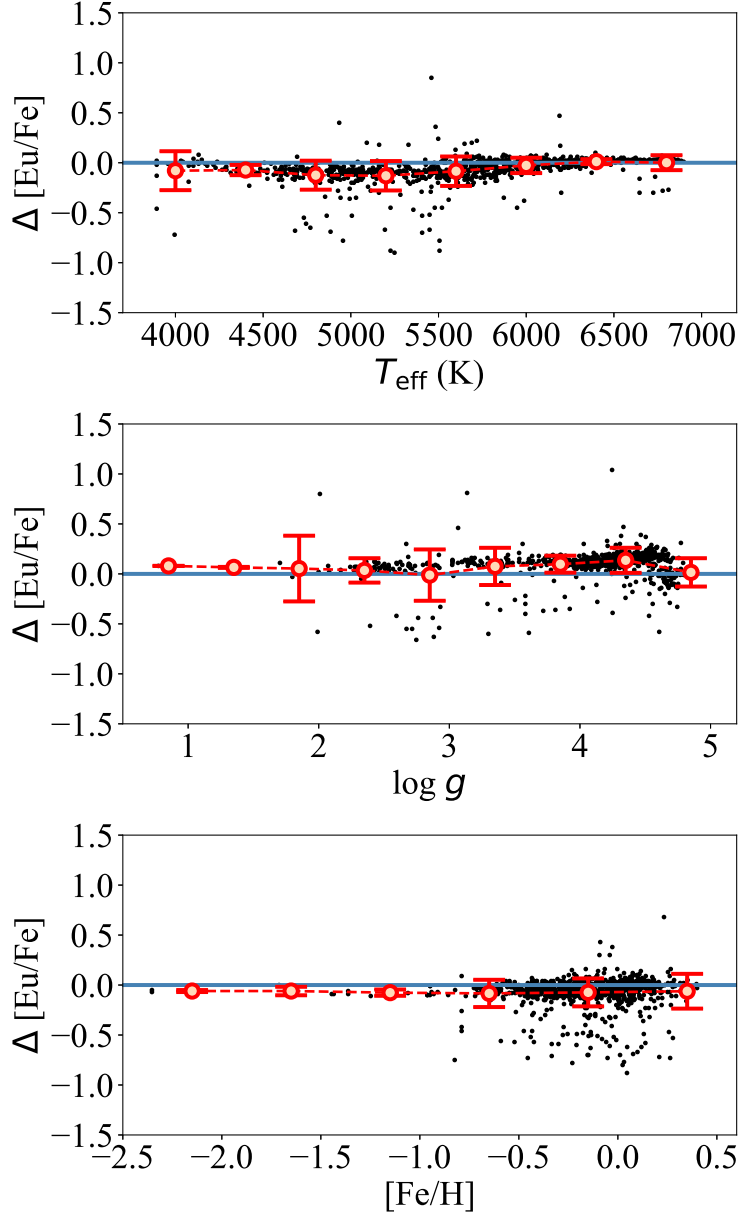}
\centering
\caption{The deviations of [Eu/Fe] obtained by varying stellar parameters with {$\mathrm{\Delta\emph{T}_{eff}}$} = $+$100.0 \,K, $\Delta$log \emph{g} = $+$0.20\,\ {dex} and $\Delta$[Fe/H] = $+$0.10 \,\ {dex} versus their corresponding changed parameters, separately in the top, middle, and bottom panels. Only results from spectra with S/N $>$ 50 are included. The sizes of the statistical bins are 400\,K, 0.5\,\ {dex} and 0.5\,\ {dex}, respectively. The average values are marked by hollow red circles while the standard deviations are given by the error bar.\label{fig:6}}
\end{figure}

\hspace{15pt}Since the stellar atmospheric parameters we adopted are from the LAMOST DR5 LRS, the typical precisions are 100\,K, 0.25\,\ {dex}, and 0.10\,\ {dex} for {$\mathrm{\emph{T}_{eff}}$}, log \emph{g} and [Fe/H], according to \cite{Luo2015}. In order to test the sensitivity of the estimated [Eu/Fe] to each parameter, we rederived the Eu abundances by increasing the temperature of $+$100\,K, surface gravity by $+$0.2\,\ {dex} and metallicity by $+$0.1\,\ {dex}, respectively. Variations of the derived [Eu/Fe] for each of these perturbations in the parameters are 
shown in the three panels of Figure \ref{fig:6}. To reduce the scattering induced by the low-S/N spectra, we only include those with S/N$>$50.  

From inspection of this figure, [Eu/Fe] may be under-estimated by about 0.1 to 0.3\,\ {dex} for stars with temperatures from 4500 to 5500\,K by increasing the effective temperature ({$\mathrm{\emph{T}_{eff}}$}) by 100\,K. While increasing the surface gravity by 0.2\,\ {dex} will lead to over-estimation of [Eu/Fe] by around 0.1 to 0.2\,\ {dex}, dwarfs tend to be more sensitive to the change of the surface gravity. It is noted that an increase of the metallicity by 0.1\,\ {dex} results in 0.1\,\ {dex} lower [Eu/Fe] abundance. A larger scatter can be found for the metal-rich objects, because the Eu II line is very weak, while the nearby Ni I line becomes very strong. Our results suggest that there are higher error rates among the RPE candidates at higher metallicities. 

We point out that the estimated stellar atmospheric parameters from the LASP pipeline, and applied to the LRS, are based on the ELODIE empirical spectral library.  However, our estimates of [Eu/Fe] and [Ba/Fe] are obtained from synthetic templates generated by SPECTRUM, which is based on the Kurucz ODFNEW atmospheres. This slight inconsistency is not expected to introduce any large offsets, as demonstrated by our comparison with the high-resolution analyses of a few of our stars by \citet{Sakari2018} and \citet{Rasmussen2020}. 


\section{Summary}
\label{sec:sum}
\hspace{15pt}We have developed an approach to search for RPE stars based on matching  LAMOST MRS spectra with synthetic templates having corresponding stellar parameters. From 12,209 objects of 32,774 spectra (S/N $>$ 50), 13 metal-poor stars with $-2.35<\rm{[Fe/H]}<-0.91$ fulfill the selection criteria. We plan to obtain high-resolution follow-up spectroscopy of these candidates, in order to validate them as bona-fide RPE stars. 

The uncertainties of the [Eu/Fe] measurements introduced by the quality of the spectra and the precisions of the adopted stellar parameters are estimated to be ~0.2\,\ {dex}. Considering this, our selection threshold is set to [Eu/Fe] $\textgreater+$0.5, which is 0.2\,\ {dex} higher than the threshold for $r$-I stars, [Eu/Fe] $\textgreater+$0.3. We note that high S/N ratios  ($\textgreater$50)  are required for reliable  identification of RPE stars. 

In a subsequent paper, we plan to search for  very metal-poor RPE candidates  ([Fe/H] $< -2$) using the improved stellar atmospheric parameters from \citet{Li2018}, the refined estimates for these stars by Beers listed in the appendix of \citet{Yuan2020}, and refined estimates for DR5 LRS very metal-poor stars presently being obtained by Beers. We note that the Ba II line at 6496.9\,\AA\ can be detected even for very metal-poor stars, while it is often difficult to detect the Eu II line at 6645.1\,\AA. Thus, we will also  include the [Ba/Fe] abundaces during this search.

\normalem
\begin{acknowledgements}
Our research is supported by National Key R\&D Program of China No.2019YFA0405502, the Key Research Program of the Chinese Academy of Sciences under grant No.XDPB09-02, the National Natural Science Foundation of China under grant Nos. 11833006, 11473033, 11603037, 11973049,11973052, and the International partnership program's Key foreign cooperation project, Bureau of International Cooperation, Chinese Academy of Sciences under grant No.114A32KYSB20160049, the Strategic Priority Research Program of Chinese Academy of Sciences, Grant No. XDB34020205. This work is supported by the Astronomical Big Data Joint Research Center, co-founded by the National Astronomical Observatories, Chinese Academy of Sciences and Alibaba Cloud. T. C. B. acknowledges partial support for this work from grant PHY 14-30152; Physics Frontier Center/JINA Center for the Evolution of the Element (JINA-CEE), awarded by the US National Science Foundation. T. C. B. is also grateful for support received through a PIFI Distinguished Scientist sward from the Chinese Academy of Sciences, which enabled a visit to China in 2019, where initial discussions of this project took place. H.-L.Y. acknowledges the supports from Youth Innovation Promotion Association, Chinese Academy of Sciences.This work is also partially supported by the Open Project Program of the Key Laboratory of Optical Astronomy, National Astronomical Observatories, Chinese Academy of Sciences.  Guoshoujing Telescope (the Large Sky Area Multi-Object Fiber Spectroscopic Telescope LAMOST) is a National Major Scientific Project built by the Chinese Academy of Sciences. Funding for the project has been provided by the National Development and Reform Commission. LAMOST is operated and managed by the National Astronomical Observatories, Chinese Academy of Sciences. 
\end{acknowledgements}

 


\bibliographystyle{raa}
\bibliography{ms2020-0236}

\end{document}